\documentclass[twocolumn,twocolappendix]{aastex631}
\usepackage{graphicx}
\usepackage{dcolumn}

\usepackage{xcolor}
\usepackage{color}
\usepackage{bm}
\usepackage{amssymb, amsmath}

\usepackage{soul}

\usepackage{cancel}

\begin{document}

\preprint{XXX}

\title{Long-lived Equilibria in Kinetic Astrophysical Plasma Turbulence}

\author[0000-0002-7459-5735]{Mario Imbrogno}
\affiliation{Dipartimento di Fisica, Universit\`a della Calabria, Arcavacata (Cosenza), 87036, IT}

\author[0000-0001-8694-3058]{Claudio Meringolo}
\affiliation{Institut für Theoretische Physik, Goethe Universität, Max-von-Laue-Str. 1, D-60438 Frankfurt am Main, DE}

\author[0000-0002-0786-7307]{Sergio Servidio}
\affiliation{Dipartimento di Fisica, Universit\`a della Calabria, Arcavacata di Rende, 87036, IT}

\author[0000-0002-3945-6342]{Alejandro Cruz-Osorio}
\affiliation{Instituto de Astronom\'{\i}a, Universidad Nacional Aut\'onoma de M\'exico, AP 70-264, 04510 Ciudad de M\'exico, MX}

\author[0000-0001-6295-596X]{Beno\^it Cerutti}
\affiliation{Univ. Grenoble Alpes, CNRS, IPAG, 38000 Grenoble, FR}

\author[0000-0002-7216-5491]{Francesco Pegoraro}
\affiliation{Dipartimento di Fisica “Enrico Fermi”, Universit\`a di Pisa, Pisa, 56122, IT}
\affiliation{INAF-IAPS, via Fosso del Cavaliere 100, Roma, {00133}, IT}



\begin{abstract}

Turbulence in classical fluids is characterized by persistent structures that emerge from the chaotic landscape. We investigate the analogous process in fully kinetic plasma turbulence by using high-resolution, direct numerical simulations in two spatial dimensions. We observe the formation of long-lived vortices with a profile typical of macroscopic, magnetically dominated force-free states. Inspired by the Harris pinch model for inhomogeneous equilibria, we describe these metastable solutions with a self-consistent kinetic model in a cylindrical coordinate system centered on a representative vortex, starting from an explicit form of the particle velocity distribution function. Such new equilibria can be simplified to a Gold-Hoyle solution of the modified force-free state. Turbulence is mediated by the long-lived structures, accompanied by transients in which such vortices merge and form self-similarly new metastable equilibria. This process can be relevant to the comprehension of various astrophysical phenomena, going from the formation of plasmoids in the vicinity of massive compact objects to the emergence of coherent structures in the heliosphere.

\end{abstract}



\section{Introduction} \label{sec:intro}

Astrophysical turbulence remains among the most fascinating phenomena, characterizing diverse systems, ranging from the heliosphere to interstellar medium and compact object environments \citep{GoldsteinEA95, Baiotti2016, RipperdaEA22}. This process covers a wide range of length- and timescales, from large-scale eddies to sub-electron scales \citep{SahraouiEA09}. In this scenario, the cascade process is envisioned as a flux of energy from large-scale shears and boundary layers to the scales typical of particle interactions, where energy conversion is taking place \citep{MatthaeusEA15}. 

Turbulence is generally synonymous with randomness and unpredictability, although this is not quite correct. Persistent, long-lived structures indeed emerge from such a chaotic state, as observed in several systems \citep{ChavanisEA98, AlexandrovaEA08, KarimabadiEA13}. This ``zoo'' of coherent patterns can be qualitatively cataloged between vortices, waves, and discontinuous layers \citep{MatthaeusEA15}. In this context, it is also important to mention large-scale structures such as Alfv\'en vortices \citep{PokhotelovEA92, AlexandrovaEA08}.

Although considerable effort has been devoted to the characterization of persistent structures in classical (viscous) fluids \citep{MontgomeryEA92}, much less is known about the collisionless, magnetized counterpart. In plasma turbulence, stable structures may permeate the system and travel undisturbed over long timescales, as one would expect in stellar winds \citep{BorovskyEA08, PecoraEA19} and accretion flows \citep{RipperdaEA20, NathanailEA22}. These patterns, known as ``plasmoids'' or ``magnetic vortices'' for their geometrical resemblance with hydrodynamical vortices, might be a crucial element of particle energization and dissipation \citep{DrakeEA10, PetropoulouEA16, KhabarovaEA21, MellahEA22}. Despite intense investigations, very little is known about their internal structure \citep{AllansonEA16, LukinEA18}, mostly because of the coupling between large scales and characteristic plasma length-scales.

In this \textit{Letter}, we describe the process of coherent structures formation in fully kinetic plasma turbulence by using numerical simulations in 2.5D (2D in space, with 3D field components). We observe the formation of long-lived coherent structures typical of macroscopic, magnetically dominated force-free states. These metastable solutions can be described with a self-consistent kinetic model, in a cylindrical coordinate frame, starting from an explicit form of the particle velocity distribution function. Such new equilibria can be simplified to a modified force-free state whose description can have several applications in all those studies concerned with the formation of coherent structures in astrophysical plasmas, such as plasmoids in accretion flows and persistent flux ropes in the solar wind.

\section{Methods} \label{sec:methods}

Our simulations are based on a full kinetic model of relativistic plasma, by using the well-tested particle-in-cell (PIC) code \texttt{Zeltron} \citep{CeruttiEA13}, which solves the equations of motion for a distribution of charged particles (i.e. characteristic curves for the Vlasov equation) when coupled to Maxwell's equations expressed in terms of a total magnetic field $\bm{b}$, an electric field $\bm{\epsilon}$, a current density $\bm{j} := \sum_{\alpha} q_{\alpha} n_{\alpha} \bm{u}_{\alpha}$, and a charge density $\rho_c := \sum_{\alpha} q_{\alpha} n_{\alpha}$, with $n_{\alpha}$ the number density of each species, ${\bm u}_{\alpha}$ the bulk velocity, $q_{\alpha}$ the charge, and $\alpha$ the species index representing either protons ($p$) or electrons ($e$).

The simulation setup follows closely that presented in \citet{MeringoloEA23}, with $N_x = N_y = 16384$ mesh points, in a square of side $L_{0} \approx 5461 ~ d_e$, being $d_e$ the electron skin depth. We employ a realistic mass ratio $m_p/m_e = 1836$ and a total number of $\approx 2.7 \times 10^9$ macro-particles. Specific details on the simulation and the parameters are given in the Appendix \ref{sec:numerical}. We impose large-scale, random initial conditions for the magnetic field with a superposition of low wavenumber Fourier modes so as to achieve a strong turbulence state \citep{MeringoloEA23}. In particular, we set $\delta b/b_{0z} \sim 1$, where $\delta b$ is the root mean square (rms) of the magnetic field fluctuations and $b_{0z}$ is the out-of-plane (along $z$) mean magnetic field strength. Given the size of the magnetization $\sigma \approx 1$, which expresses the ratio between the magnetic pressure and the enthalpy density (the latter is the sum of the mass-energy density $\sum_{\alpha} n_{\alpha} m_{\alpha} c^2$ and the internal energy density $\sum_{\alpha} \varrho_{\alpha} \approx \sum_{\alpha} P_{\alpha}(\Gamma_{\alpha} - 1)^{-1}$ weighted by the adiabatic index $\Gamma_{\alpha}$, where $P_{\alpha}$ is the partial pressure related to the $\alpha$-th species), and the plasma beta $\beta_p = \beta_e = 3 \times 10^{-3}$, which quantifies the ratio between kinetic and magnetic pressure, the observed dynamics is that of a weakly relativistic plasma. \\

\begin{figure}[b]
	\centering
	\includegraphics[width=0.99\columnwidth]{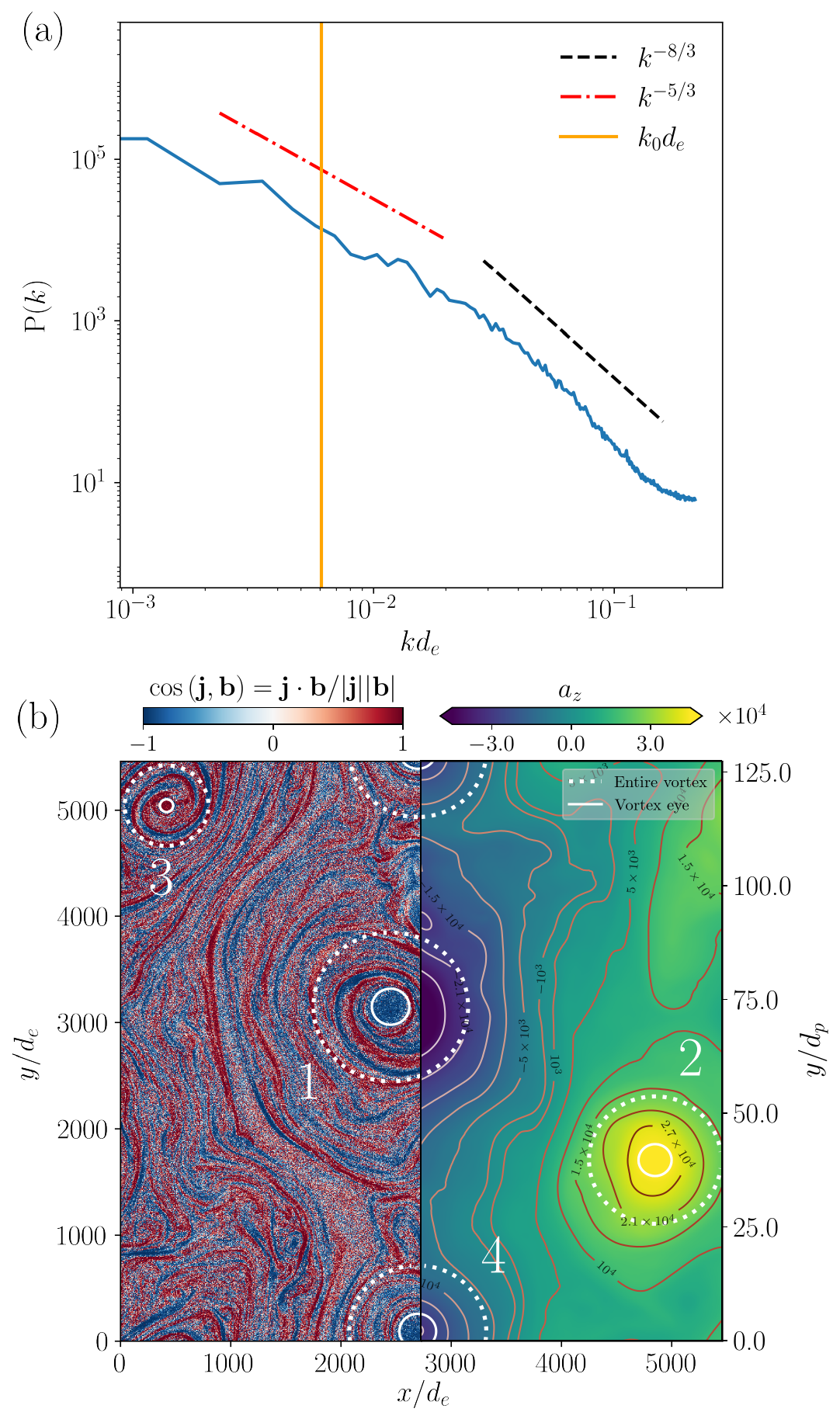}
	\caption{(a) Magnetic power spectrum at a time beyond the peak of nonlinear activity vs $k d_e$. The vertical (orange) line represents the persistent structures wavenumber $k_0 d_e$, which identifies the typical scale of the vortices. (b) 2D contour of the magnetic vector potential component $a_z$ (right side) and the cosine angle between the current density and the total magnetic field (left side). The y-axis is normalized to both the electron (left side) and the proton skin depth $d_p$ (right side).}
	\label{fig:speccontours}
\end{figure}

\section{Turbulence and coherent structures} \label{sec:turbulence}

The initial conditions rapidly produce a turbulent cascade, similar to fluid plasma models \citep{ServidioEA09}. We concentrate our analysis when the peak of the averaged current density, which also corresponds to the time of most intense nonlinear activity \citep{ServidioEA12}, has long been reached (see the end of Appendix \ref{sec:numerical} for further information). Furthermore, under these conditions, a balance between the large-scale energy flux and the collisionless energy-conversion mechanisms is established, thus yielding a quasi-steady state. The power spectrum at that time is fully developed, as reported in Fig.~\ref{fig:speccontours}-(a), and is consistent with typical observations of astrophysical turbulence \citep{BaleEA05, SahraouiEA09, AlexandrovaEA09}.

As the broadband turbulence develops, we observe the appearance of coherent structures that move through the turbulent background, as can be inferred from the out-of-plane component of the magnetic potential $a_z$ in Fig.~\ref{fig:speccontours}-(b). Similarly to MHD \citep{PouquetEA08, ServidioEA08}, kinetic turbulence tends to form local correlations, where the current $\bm{j}$ manifests a substantial tendency to align with the total magnetic field $\bm{b}$. We estimate its strength by computing the $\cos({\bm j}, {\bm b}) = {\bm j}\cdot {\bm b}/|{\bm j}||{\bm b}|$, and as displayed in Fig.\ref{fig:speccontours}-(b), dominant structures are visible, whose morphology resembles hydrodynamic swirls and cyclones. They have a clear ``eye", where the alignment is net, and an outside region with advecting arms, where the cosine regularly changes sign while maintaining an overall circular symmetry. The characteristic size of such structures is $\approx 140 ~ d_e (3 ~ d_p)$ for the eye and $\approx 515 ~ d_e (12 ~ d_p)$ for the spiral arms, typical of inertial range turbulence, as can be seen from Fig.~\ref{fig:speccontours}-(a). Note that the largest vortices (eye and arms) are on the order of a few $d_p$,  corresponding to the correlation length, $\lambda_C$, set by our initial conditions of homogeneous turbulence, which, for the analyzed configuration, is $\lambda_C \approx 440 \, d_e \approx 10 \, d_p$.

To gain insight into the properties of these long-lived features, we focus our attention on regions where the magnitude of the magnetic potential $a_z$ at the center of the structure is much larger than its own rms \citep{ServidioEA09}. After the time of maximum turbulence, we select four main vortices, which are marked with (white) circumferences in Fig.~\ref{fig:speccontours}-(b). For each of these, we consider a local cylindrical coordinate system centered at the O-point of the vector potential and use it to produce locally azimuthally averaged quantities. More precisely, for any generic field $h$, we first transform it to the cylindrical coordinate frame, i.e. $h(x, y) \to h(r, \phi)$, and then compute the corresponding average as $H(r) := ({2\pi})^{-1} \int_0^{2\pi} h(r, \phi^\prime) \, d\phi^\prime$ (hereafter, we use capital letters to indicate azimuthally averaged quantities).

\begin{figure}[ht]
	\centering
	\includegraphics[width=0.99\columnwidth]{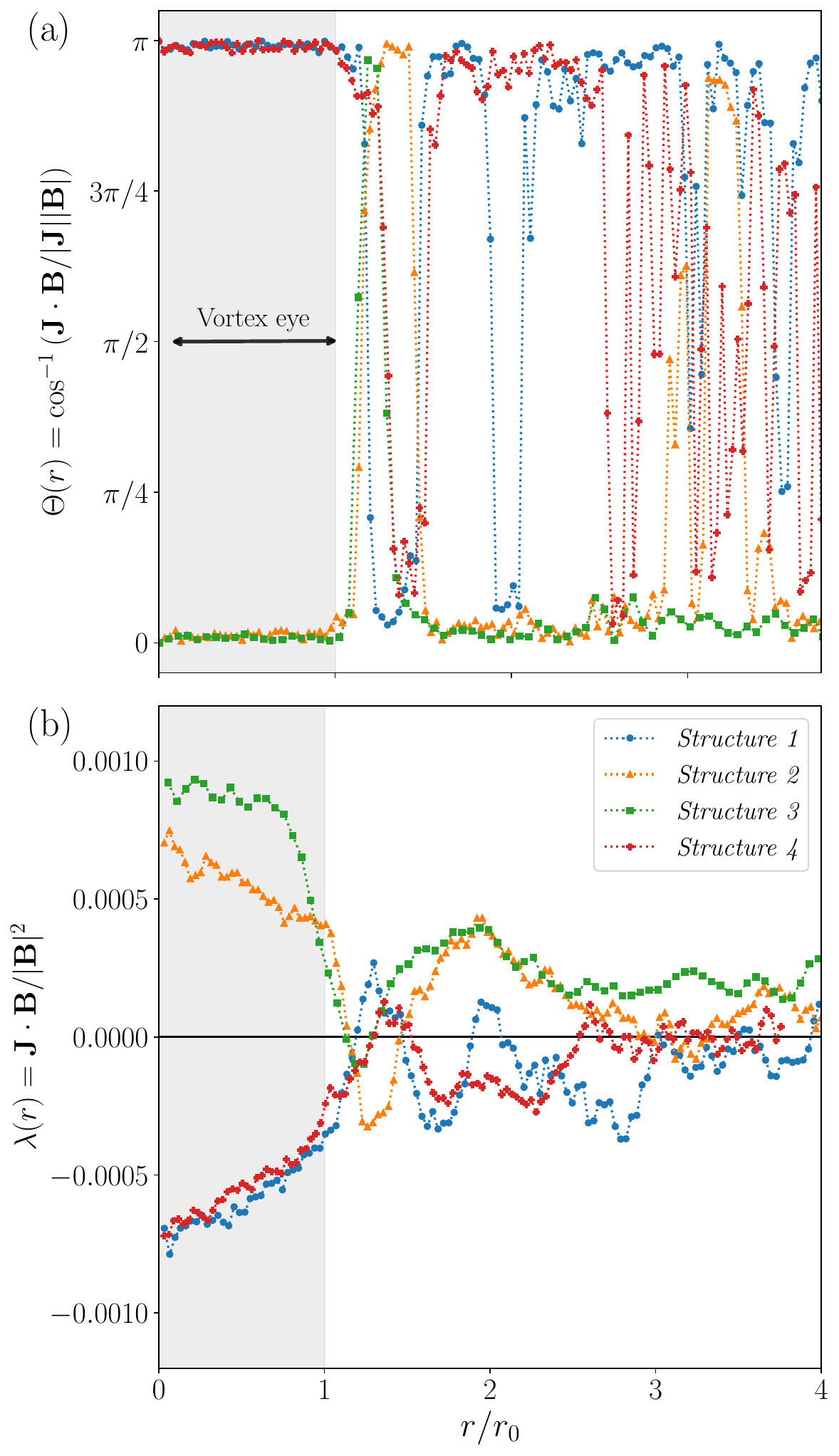}
	\caption{(a) Radial behavior of the angle between the current density and the total magnetic field (azimuthal averages) for each structure depicted in Fig.~\ref{fig:speccontours}-(b). (b) Force-free parameter $\lambda(r)$. We define $r_0$ as the vortex eye radius, namely the position of the first spiraling arm.}
	\label{fig:anglelambda}
\end{figure}

In Fig.~\ref{fig:anglelambda}-(a), we show the angle between the azimuthally-averaged current density $\bm J$ and the magnetic field $\bm B$, namely $\Theta = \cos^{-1}{ \left( {\bm J}\cdot{\bm B}/|{\bm J}||{\bm B}| \right)}$, evaluated inside the structures. The alignment is here extremely high and all the vortices, which vary in size and intensity, exhibit the same behavior: they form highly aligned cores (eyes) and wall boundaries followed by spiraling arms \citep{servidio2010local, mcwilliams1984emergence, powell1998surface,  haller2005objective, carnevale1991evolution}. Such spirals manifest alternate ${\bm J} - {\bm B}$ alignment, implying a characteristic radial mode. We denote $r_0$ as the radius of the vortex eye, defined as the distance from the center to the first spiraling arm where the scalar product $\bm{J} \cdot \bm{B}$ changes sign (see Fig.~\ref{fig:anglelambda}). The alignment is progressively weakened, and the plasmoid then blends with the background on scales $\mathcal{O}(4 r_{0})$, which qualitatively coincide with the last closed magnetic surfaces [dotted lines in Fig.~\ref{fig:speccontours}-(b)].

The above alignment suggests the tendency of the system to produce force-free states, where $\bm{J} = (4 \pi)^{-1} \bm {\nabla} \times \bm{B} = \lambda \bm{B}$, as in large-scale fluid models \citep{taylor1974relaxation}. In this regard, we show the force-free parameter $\lambda(r) = {\bm J} \cdot {\bm B}/B^2$ in Fig.~\ref{fig:anglelambda}-(b).	In contrast to the classical and global (constant-$\lambda$) force-free states, our local version of the minimization process (for each vortex) reveals a strong dependency of $\lambda$ as a function of $r$, approaching zero at the vortex boundaries -- namely a nonlinear force-free state with uniform twist per unit length. This radial dependency of $\lambda$ might suggest a more complex relaxation process \citep{MontgomeryEA92, ServidioEA10b}.

\section{A kinetic model} \label{sec:kinmodel}

In what follows, we discuss how to interpret the long-lived structures in terms of kinetic plasma theory. We recall that the Vlasov equation for the $\alpha$-th species distribution function $f_\alpha({\bm x}, {\bm v}, t)$ can be written as $\partial f_{\alpha}/\partial t = \{ H, f_{\alpha} \}$, where $H$ is the particle Hamiltonian and $\{ \cdot , \cdot\}$ are the standard Poisson brackets. A stationary equilibrium is thus characterized by $ \{ H, f_{\alpha} \} = 0$, where $f_{\alpha}$ must be represented as a function of the integrals of motion. Inspired by the popular Harris approach \citep{Harris62}, we consider an exponential dependence on the invariants (energy and momenta), neglecting relativistic corrections (the bulk flows are nonrelativistic), with a simple drifting-Maxwellian in a cylindrical coordinate system 
\begin{equation}
	\! f_{\alpha}(\bm{x}, \bm{v}) \! = \! f_{\alpha 0}
	\exp{ \left[ \! - \dfrac{ \mathcal{E}_{\alpha} \! - \! v_{\alpha}^* P_{z \alpha} \! - \! \Omega_{\alpha}^* P_{\phi \alpha}}{k_B T_{\alpha}} \! \right] }, \label{eq:equilibrium1}
\end{equation}
where $f_{\alpha 0} := N_{\alpha 0} \left( m_{\alpha}/2 \pi k_B T_{\alpha} \right)^{3/2}$ is a normalization constant, $T_\alpha$ the temperature, $\mathcal{E}_{\alpha} := m_{\alpha} (v_r^2 + v_{\phi}^2 + v_z^2) /2 + q_{\alpha} \psi(r)$ the particle energy, $P_{\phi\alpha} := r \left[ m_{\alpha} v_{\phi} + q_{\alpha} A_\phi(r)/c \right]$ and $P_{z \alpha} := m_{\alpha} v_{z} + q_{\alpha} A_z(r)/c$ the azimuthal and vertical momentum, respectively ($A_i$ are the components of the averaged vector potential in the Lorenz gauge, $\psi$ is the electrostatic potential, and $v_i$ are the components of the particle velocities). As usual in literature, $k_B$ is the Boltzmann constant, $c$ the speed of light, and $m_{\alpha}$ the rest mass of each species. The undetermined (free) quantities appearing in Eq.~(\ref{eq:equilibrium1}) stand for the out-of-plane characteristic linear velocity $v_{\alpha}^*$ and azimuthal velocity $\Omega_{\alpha}^*$. Once these quantities are specified, an exact kinetic equilibrium can be constructed.

\begin{figure}[ht]
	\centering
	\includegraphics[width=0.99\columnwidth]{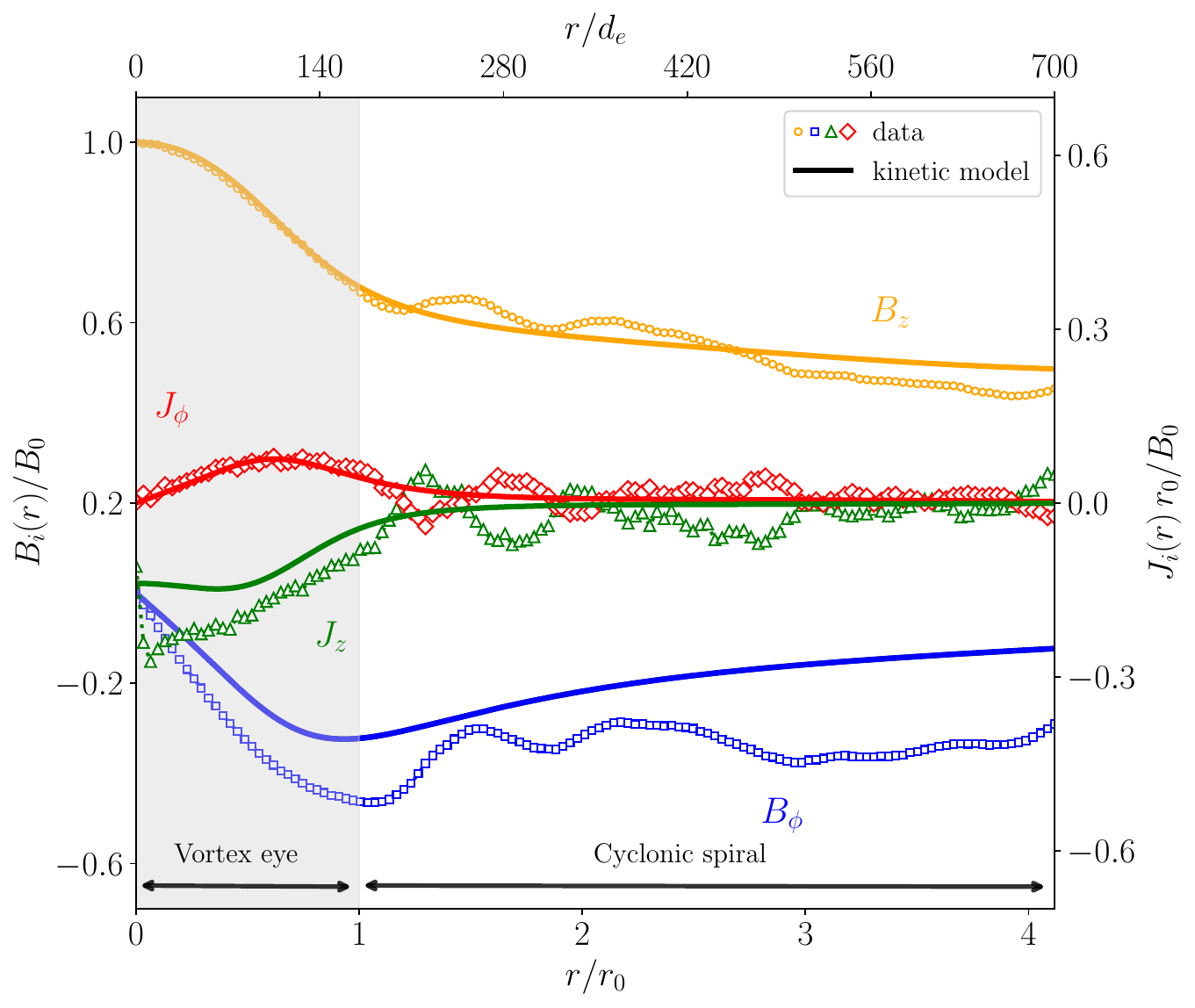}
	\caption{(a) Total magnetic field and current density components for the central vortex in Fig.~\ref{fig:speccontours} (open symbols) and the KVR model (solid lines). Data have been averaged over time in the vortex frame.}
	\label{fig:modelkvr}
\end{figure}

By taking the moments of Eq.~(\ref{eq:equilibrium1}), it is possible compute the particle number densities $N_{\alpha}(r) = \int f_{\alpha} d^3 \bm{v}= \exp{ \left[ \gamma_{\alpha}(r) \right] }$, where
\begin{equation}
	\begin{aligned}
		\gamma_{\alpha}(r) \! & = \! \ln{(N_{0 \alpha})} \! + \! \dfrac{q_{\alpha}}{c k_B T_{\alpha}} \bigg[ v_{\alpha}^{*} A_z(r) \! + \! r \Omega_{\alpha}^{*} A_{\phi}(r) \\
		& - \! c \psi(r) \bigg] \! + \! \dfrac{m_{\alpha}}{2 k_B T_{\alpha}} \left( r^2 \Omega_{\alpha}^{{*}^2} \! + \! v_{\alpha}^{{*}^2} \right).
	\end{aligned}
	\label{eq:n}
\end{equation}
Similarly, by computing the azimuthally-averaged bulk velocity ${\bm U}(r)= N^{-1}(r) \int {\bm v} f d^3v$, it is simple to demonstrate that it is related to the free parameters, being $U_{\alpha z}(r) = v_{\alpha z}^*$ and $U_{\alpha \phi}(r) = \Omega_{\alpha}^* r$. The problem can be further simplified with an assumption similar to the one made within the Harris sheet pinch, i.e., by imposing a (negligible) net constant charge density $\kappa := \exp{ \left[ \gamma_e(r) \right]} - \exp{ \left[ \gamma_p(r) \right]}$ that yields
\begin{equation}
	\begin{aligned}
		& v_p^{*} \! = \! - \dfrac{T_p}{T_e} v_e^{*}, ~~ \Omega_p^{*} \! = \! - \dfrac{T_p}{T_e} \Omega_e^{*}, \\
		& \kappa \! = \! \frac{ m_p {\Omega_p^{*}}^2 T_e - m_e {\Omega_e^{*}}^2 T_p}{2 \pi e^2 (T_p+T_e)}.
	\end{aligned}
	\label{eq:vioi}
\end{equation}
By using Eq.~(\ref{eq:equilibrium1}) into Maxwell's equations via the densities and the current expressions, assuming stationarity ($\partial/\partial t = 0$), and imposing the simplifications (\ref{eq:vioi}), it is possible to obtain a set of ODEs constituting our Kinetic Vortex Reconstruction (KVR) model:
\begin{eqnarray}
	&& \!\!\!\!\!\!\!\!\!\!\!\! \dfrac{d A_z}{d r} \! = \! - B_{\phi}(r)\,, \label{eq:az} \\
	&& \!\!\!\!\!\!\!\!\!\!\!\! \dfrac{d A_{\phi}}{d r} \! = \! - \dfrac{A_{\phi}(r)}{r} \! + \! B'_z(r)\,, \label{eq:aphi} \\
	&& \!\!\!\!\!\!\!\!\!\!\!\! \dfrac{d B'_z}{d r} \! = \! \dfrac{4 \pi e}{c}\left( \Omega_e^{*} \! - \! \Omega_p^{*}\right) r \, e^{\gamma_e(r)} \! + \! \Omega_p^{*} \kappa r\,, \label{eq:bz} \\
	&& \!\!\!\!\!\!\!\!\!\!\!\! \dfrac{d B_{\phi}}{d r} \! = \! -
	\dfrac{B_{\phi}(r)}{r} \! + \! \dfrac{4 \pi e}{c} \left( v_p^{*} \! -
	\! v_e^{*} \right)e^{\gamma_e(r)}\! - \! v_p^{*}
	\kappa\,. \label{eq:bphi}
\end{eqnarray}
Here, $B'_z$ denotes the fluctuations in the out-of-plane field, i.e., $B_z = B'_z + b_{0z}$. Eq.s~(\ref{eq:az})--(\ref{eq:bphi}) can be integrated numerically after specifying, for each vortex, the internal temperatures of the eye $T_\alpha$, and the free parameters $(v_{e}^*$, ~ $\Omega_{e}^*)$. The latter are obtained through a data-driven Monte Carlo method. A discussion on the reconstruction process can be found in the Appendices \ref{sec:kineq} and \ref{sec:data-driven}. It should be emphasized that a charge separation, albeit small (less than $10 \, \%$), is present, especially in the vortex eye (not shown here).

Fig.~\ref{fig:modelkvr} offers a direct comparison between the KVR model (lines) and the actual numerical data (symbols). The model captures the behavior of all the fields and does so particularly well in the inner regions of the vortex ($r \lesssim r_0$) and for the magnetic field dependence. Note the presence of an azimuthal current $J_{\phi}$ vanishing at the center and asymptotically, and a non-zero vertical current $J_z$ at the axis. 
\begin{figure}[b]
	\centering
	\includegraphics[width=0.99\columnwidth]{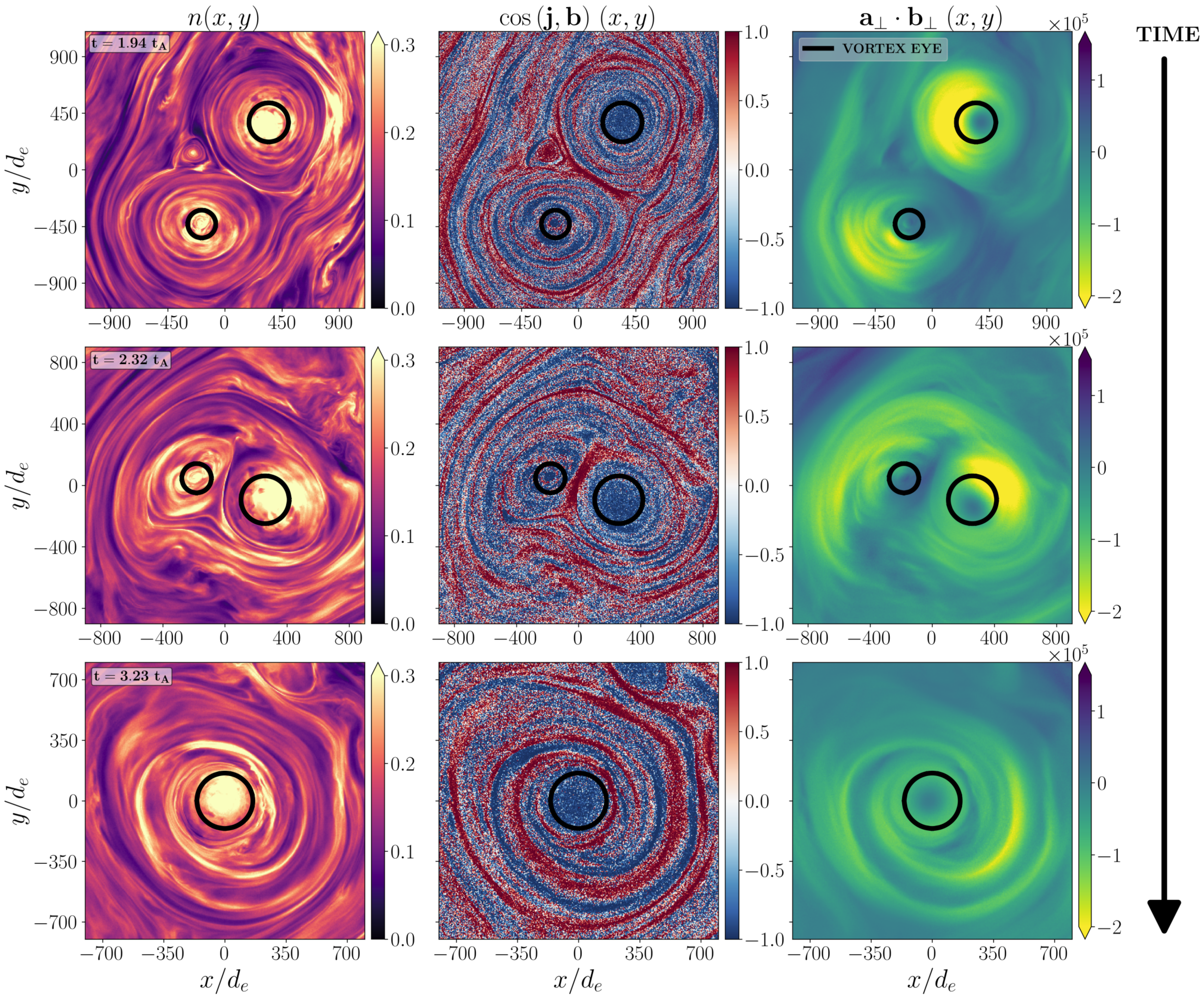}
	\caption{\textit{Merging history} of two long-lived structures as a function of Alfv\'enic crossing time (from top to bottom). The plot shows the plasma density (left column), the (force-free) cosine angle (middle column), and the magnetic helicity density (right column). Circumferences (in black) represent the eye of each vortex.}
	\label{fig:evolution}
\end{figure}
Even outside the eye, the magnetic field components are qualitatively consistent with the model, even though the data is subject to the cyclonic arms, observed in Fig.s~\ref{fig:speccontours}--\ref{fig:anglelambda}. Very similar behaviors are also shown by the other vortices tracked during the evolution: $B_z$ manifests a peak in the eye and diminishes for large $r$, whereas $B_\phi$ peaks at the vortex wall.

To understand the merging dynamics of existing vortices, we examine the simulation at different times, identifying and monitoring the most stable and long-lasting structures. In particular, we report in Fig.~\ref{fig:evolution} the \textit{``merging history''} of two individual, isolated vortices (or magnetic islands) that wander across the background until they eventually merge with a companion. The figure shows, as a function of time (from top to bottom), the plasma number density, the cosine angle, and the magnetic helicity density (see later). If the structures are both very energetic, the encounter is quite explosive, resulting in a net current layer between them, where magnetic reconnection occurs \citep{RipperdaEA19, ServidioEA09} and non-thermal particles are produced \citep{ComissoEA19}. The $\cos(\bm{j}, ~ \bm{b})$ is quite strong inside the structures, and it changes sign in between them. The example reported in Fig.~\ref{fig:evolution} also helps us to appreciate that the spiral arms described above are actually the heritage of merger events. The process is related to the conservation of magnetic helicity, in a picture similar to \citet{AlexakisEA06}, where the magnetic helicity inverse cascade can occur at all scales. In our case, this results in meta-stable vortices ranging from large injection scales (a few $d_p$) to electron scales, within the sub-inertial range.

We now reconcile the KVR model in Fig.~\ref{fig:modelkvr} with the fluid-like states in Fig.~\ref{fig:anglelambda}. By taking the moments of the Vlasov equation, it is possible to define a hierarchy of the different contributions intervening in determining the equilibrium in the islands and conclude that the magnetic forces dominate the dynamics of the long-lived equilibria (see Appendix \ref{sec:momentum} for details). As a result, neglecting secondary effects such as the charge separation, one can combine Eq.s~(\ref{eq:bz})--(\ref{eq:bphi}) as
\begin{equation}
	{\bm \nabla} \times {\bm B} \! = \! \frac{4 \pi}{c} {\bm J}_0^{*}e^{\gamma_e(r)}, 
	\label{eq:compact}
\end{equation}
where ${\bm J}_0^{*} := q_e\left[0, ~ \left(\Omega_p^* - \Omega_e^*\right)r, ~ \left(v_p^* - v_e^*\right)\right]$ is a current-to-number density and $\gamma_e(r)$ is defined in Eq.~(\ref{eq:n}). When $r \rightarrow 0$,  all functions, including $\gamma_e(r)$ and $\psi(r)$, are regular, smooth, and continuous (see Appendix \ref{sec:kineq}). Furthermore, for all vortices, we observe that $B_z$ peaks near the origin at a non-zero value, and the azimuthal component $B_\phi \sim \alpha r$, where $\alpha$ is a constant. Phenomenologically, to suppress the Lorentz force term in the Momentum equation, the magnetic field $\bm{B}$ tends to align with ${\bm J}\simeq{\bm J}_0^{*}e^{\gamma_e(r)}$, so Eq.~(\ref{eq:compact}) can be approximated as
\begin{equation}
{\bm \nabla} \times {\bm B} = f(r) {\bm J}_0^{*} \, \sim \, \lambda(r) {\bm B}(r),
\label{eq:approx}
\end{equation}
where all the rescaling constraints are included in $\lambda(r)$. A very robust solution to Eq.~(\ref{eq:approx}) is known as the Gold-Hoyle (GH) vortex \citep{gold1960origin} -- a flux tube describing a force-free, twisted field. This equilibrium was discovered for force-free coronal structures and has potential uses in astrophysical contexts \citep{fushiki19953}. It is given by \citep{AllansonEA16}
\begin{eqnarray}
	&& \bm{A}(\tilde{r}) \! = \! \dfrac{B_0}{2 \xi} \bigg( \! 0, \! \dfrac{1}{\tilde{r}} \ln ( 1 \! + \! \tilde{r}^2 ), \! \mp \ln \left( 1 \! + \! \tilde{r}^2 \right) \bigg), \label{eq:agh}\\
	&& \bm{B}(\tilde{r}) \! = \! B_0 \bigg( \! 0, \! \pm\dfrac{\tilde{r}}{1 \! + \! \tilde{r}^2}, \! \dfrac{1}{1 \! + \! \tilde{r}^2} \bigg), \label{eq:bgh}\\
	&& \bm{J}(\tilde{r}) \! = \! 2 \dfrac{\xi B_0}{\mu_0} \bigg( \! 0, \! \dfrac{\tilde{r}}{(1 \! + \! \tilde{r}^2)^2}, \! \pm\dfrac{1}{(1 \! + \! \tilde{r}^2)^2} \bigg), \label{eq:jgh}
\end{eqnarray}
where $B_0$ is a typical field strength, $\tilde{r} = \xi r$ is a dimensionless radial coordinate, and $\xi$ a characteristic gradient. The solution is valid for ``strong'' vortices (fluctuations of the order of the mean field) and generalized for clockward (upper sign) or anti-clockward (lower sign) rotation.

\begin{figure}[b]
	\centering
	\includegraphics[width=0.99\columnwidth]{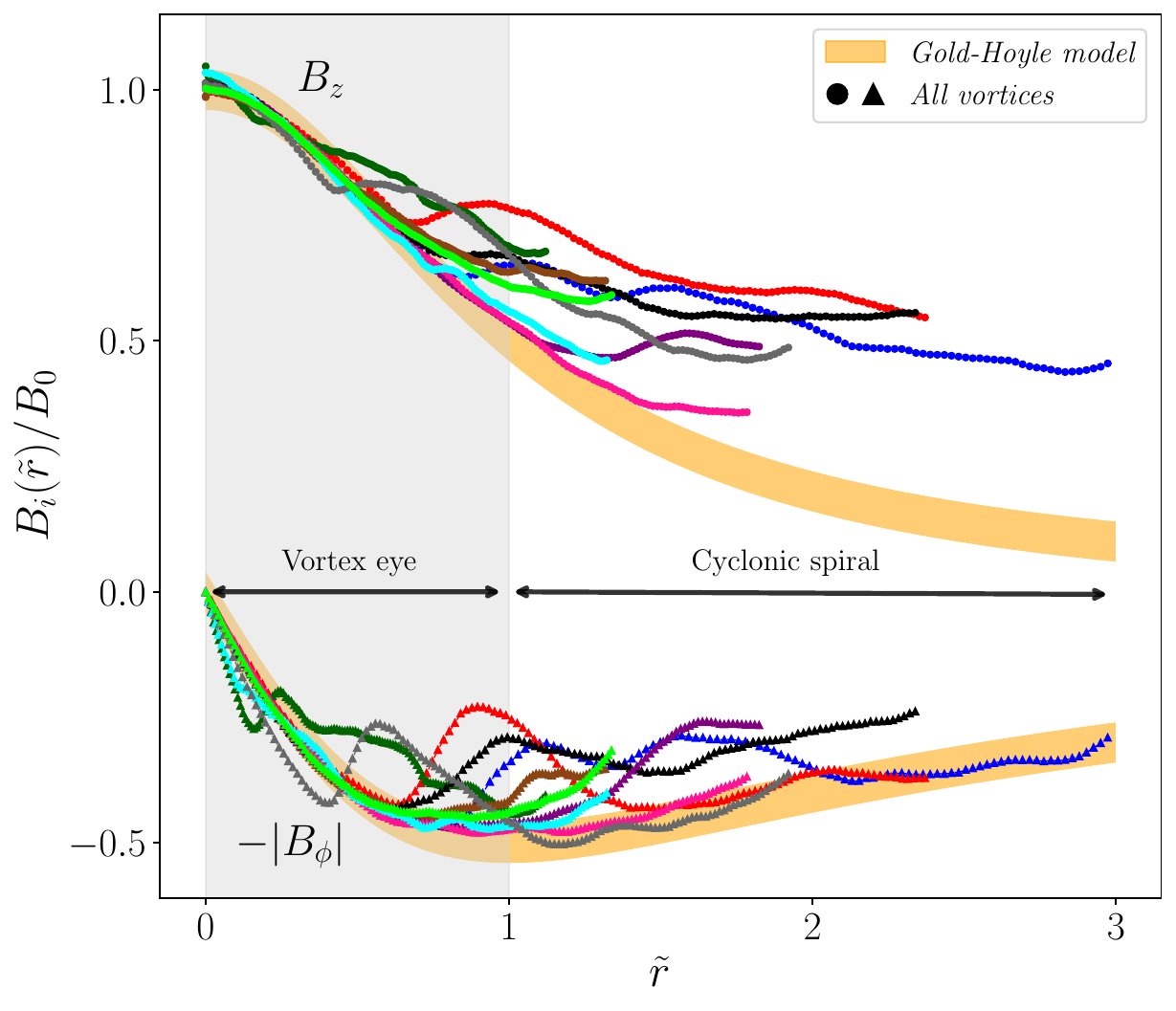}
	\caption{Radial behavior of the magnetic field components of all the long-lived vortices appearing during the simulation (points). Components have been rescaled to the field at the center $B_0$ and	using their typical gradient $\xi$. The GH solution (shaded) describes qualitatively well the profiles, near the eye.}
	\label{fig:GHall}
\end{figure}

To determine whether the above model is a universal property of the turbulent cascade, we detect and analyze all the (ten) long-lived structures appearing during the simulation, collected in Fig.~\ref{fig:GHall}. Here, the field components are normalized to $B_0$ and rescaled via $\xi$, which is determined through a simple fitting procedure; also reported are the corresponding GH magnetic components from Eq.~(\ref{eq:bgh}) (to avoid overcrowding, we represent $-|B_{\phi}|$). The GH solution provides a qualitatively accurate description, implying that the KVR (and its GH approximation) may characterize energetic structures in plasma turbulence. This kind of solution is achieved through a local relaxation process where the magnetic helicity is finite and plays a crucial role \citep{Woltjer58, taylor1974relaxation, MatthaeusEA82, AlexakisEA06}. Such an MHD invariant measures the twisting of the field lines and, in 2.5D, is defined as the volume average $H_m = \mathcal{V}^{-1} \int_{\bm{\mathcal{V}}} ~ h_m d^3x$, where $h_m = {\bm a}_\perp \cdot {\bm b}_\perp$, with ${\bm a}_\perp = (a_x, \, a_y)$ and ${\bm b}_\perp = (b_x, \, b_y)$ being the in-plane components. Contrary to the classical Bessel solutions of the linear force-free state, the numerical results are more consistent with a constant twist per unit length (along the vortex axis). As shown in Fig.~\ref{fig:evolution}, the structures retain a finite amount of magnetic helicity, before and after the merging. From Eq.s~({\ref{eq:agh})--({\ref{eq:bgh}), performing a volume average over the flux tube, one gets $H_m \sim 10^{-1}B_0^2 r_0$ (e.g., $B_0 \approx 43$ and $r_0 \approx 170 \, d_e$ for Vortex 1 in Fig.~\ref{fig:speccontours}-(b)), which is in accordance with the observed values in Fig.~\ref{fig:evolution} ($H_m \sim 3 \times 10^{4}$). It is particularly interesting to note that all the structures described here in the context of kinetic theory closely resemble those those observed in the magnetosheath \citep{AlexandrovaEA06} and in the solar wind, with sizes spanning from MHD to sub-ion scales \citep{VinogradovEA23}.

\section{Conclusions} \label{sec:conclusions}

Exploiting the results of direct numerical PIC simulations, we propose a description of plasma turbulence envisioned as a mosaic of equilibrium-like patterns. In such a scenario, coherent structures emerge from the turbulent background and occasionally encounter and merge with other similar metastable structures during their life. This self-similar process systematically produces new-born equilibria, obeying a kinetic stationary solution of the Vlasov equation. The magnetic vortices qualitatively obey a universal form that can be simply characterized using the Gold-Hoyle equilibrium. These structures show a characteristic size typical of inertial range turbulence, thus making them macroscopically relevant: they might grow, by coalescence, to a significant fraction of astrophysical system size, with potential observable signatures \citep{MellahEA22, VosEA23}. The present work focuses on a 2.5D model, considering the plane perpendicular to a mean magnetic field. While this two-dimensional approximation differs qualitatively from the (more complex and expensive) 3D case, it may still provide insights into some relaxation processes characteristic of magnetized astrophysical plasmas. Indeed, these equilibria might be relevant also for full 3D anisotropic settings, in cases where an external field effectively reduces the dimensionality of turbulence \citep{ShebalinEA83, KhabarovaEA21, RipperdaEA22, ChernoglazovEA21}. In such a general case, which will motivate future investigations, the KVR model can acquire a weak dependency along the magnetic field coordinate, say $z$, as typical of solar flux ropes. We plan to investigate the dynamics of magnetic helicity in these vortices on macroscopic scales (in a fluid-like regime) by using larger domains. Additionally, in future work, we will describe the formation process of each individual vortex and how the GH solution is achieved over time, while also exploring the plasma parameter space.

Since the long-lived structures might potentially grow, our results could be significantly relevant for the comprehension of astrophysical plasmas, especially in scenarios where transient and flare emissions are associated with the formation of plasmoids during the accretion process, as observed, for instance, in SgrA*, as well as in the observation of flux tubes in the solar wind and corona (varying the magnetization $\sigma$ and the plasma $\beta$ parameters).

\section*{Acknowledgments}
The authors thank Luciano Rezzolla and William H. Matthaeus for useful discussions. 
CM acknowledges the support from the ERC Advanced Grant ``JETSET: Launching, propagation and emission of relativistic jets from binary mergers and across mass scales'' (Grant No. 884631).
BC acknowledges the support from the European Research Council (ERC) under the European Union’s Horizon 2020 research and innovation program (Grant Agreement No. 863412). SS acknowledges ``Progetto STAR 2-PIR01 00008'' (Italian Ministry of University and Research). ACO gratefully acknowledges ``Ciencia Básica y de Frontera 2023-2024" program of the ``Consejo Nacional de Humanidades, Ciencias y Tecnología" (CONAHCYT, Mexico), projects CBF2023-2024-1102 and 257435. Computational resources were provided by CINECA through the ISCRA Class B project ``KITCOM - HP10BB7U73''. Finally, the authors would like to thank the anonymous Reviewer for the very useful comments and suggestions.

\appendix

\section{The numerical method} \label{sec:numerical}

Simulations are performed by means of the PIC code \texttt{Zeltron}, which solves the following system of equations (Lorentz--Newton $+$ Maxwell's equations) for particles and fields evolution:
\begin{eqnarray}
&& {\bm v}_i \! = \! \dfrac{d \bm{r}_i}{dt} \! = \! \dfrac{c \bm{u}_i}{W_i}, \label{eq:va} \\[2pt]
&& \dfrac{d \bm{u}_i}{dt} \! = \! \dfrac{q_\alpha}{m_\alpha c} \left[ \bm{\epsilon} \! + \! \dfrac{\bm{v}_i \times \bm{b}}{c} \right],  \label{eq:duadt} \\[2pt]
&& {\bm \nabla}\cdot {\bm \epsilon} \! = \! 4 \pi \rho_c, \label{Gauss11} \\[2pt]
&& {\bm \nabla}\cdot {\bm b} \! = \! 0, \label{Gauss21} \\[2pt]
&& \dfrac{\partial \bm{b}}{\partial t} \! = \! -c \bm{\nabla} \times \bm{\epsilon}, \label{farad1} \\[2pt]
&& \dfrac{\partial \bm{\epsilon}}{\partial t} \! = \! c \bm{\nabla} \times \bm{b} \! - \! \mu_0 \bm{j}. \label{ampere1}
\label{parmot1}
\end{eqnarray}
The first two equations describe the motion of particles in the Lagrangian specification and constitute the characteristic curves along which the Vlasov equation can be solved. Here, $\bm{r}_i$ and $\bm{v}_i$ are the position and the proper 3-velocity of the $i$-th macro-particle \citep{CeruttiEA13}, ${\bm u}_i = \bm{p}_i \, (m_{\alpha} c)^{-1}$ represents its normalized momentum as measured by an inertial observer at rest (being $\bm{p}_i$ the particle's momentum in that frame), and $W_i = (1 - v_i^2/c^2)^{-1/2}$ is the associated Lorentz factor. The subscript $\alpha$ stands for either protons ($p$) or electrons ($e$). Regarding Maxwell's equations, all quantities are well-described at the beginning of Sec. \ref{sec:methods} in the \textit{Letter}.

We adopt the geometrized unit system, where the speed of light $c$, the gravitational constant $G$, the elementary charge $q_e$, the electron mass $m_e$, the Boltzmann constant $k_B$, and the reduced Planck constant $\hbar$ are set to unity, whereas the vacuum permittivity $\varepsilon_0 = 1/ 4 \pi$ and the vacuum permeability $\mu_0 = 4 \pi$. 

The code utilizes the Yee algorithm \citep{YeeEA66} to solve the time-dependent Maxwell’s equations, in which the different components of the fields are staggered in both space (on the grid) and time. \texttt{Zeltron} has a second-order error in space and time, ensuring the magnetic constraint $\bm{\nabla} \cdot \bm{b} = 0$ to be satisfied at any timestep of the simulation. All the quantities are expressed in terms of electron skin depths, i.e., $d_e := c/\omega_{pe} = c \sqrt{m_e/4 \pi n_0 q_e^2} = 1$, being the number density at equilibrium $n_0 = n_e = n_p = (4 \pi)^{-1}$.

\begin{figure}[b]
	\centering
	\includegraphics[width=0.99\columnwidth]{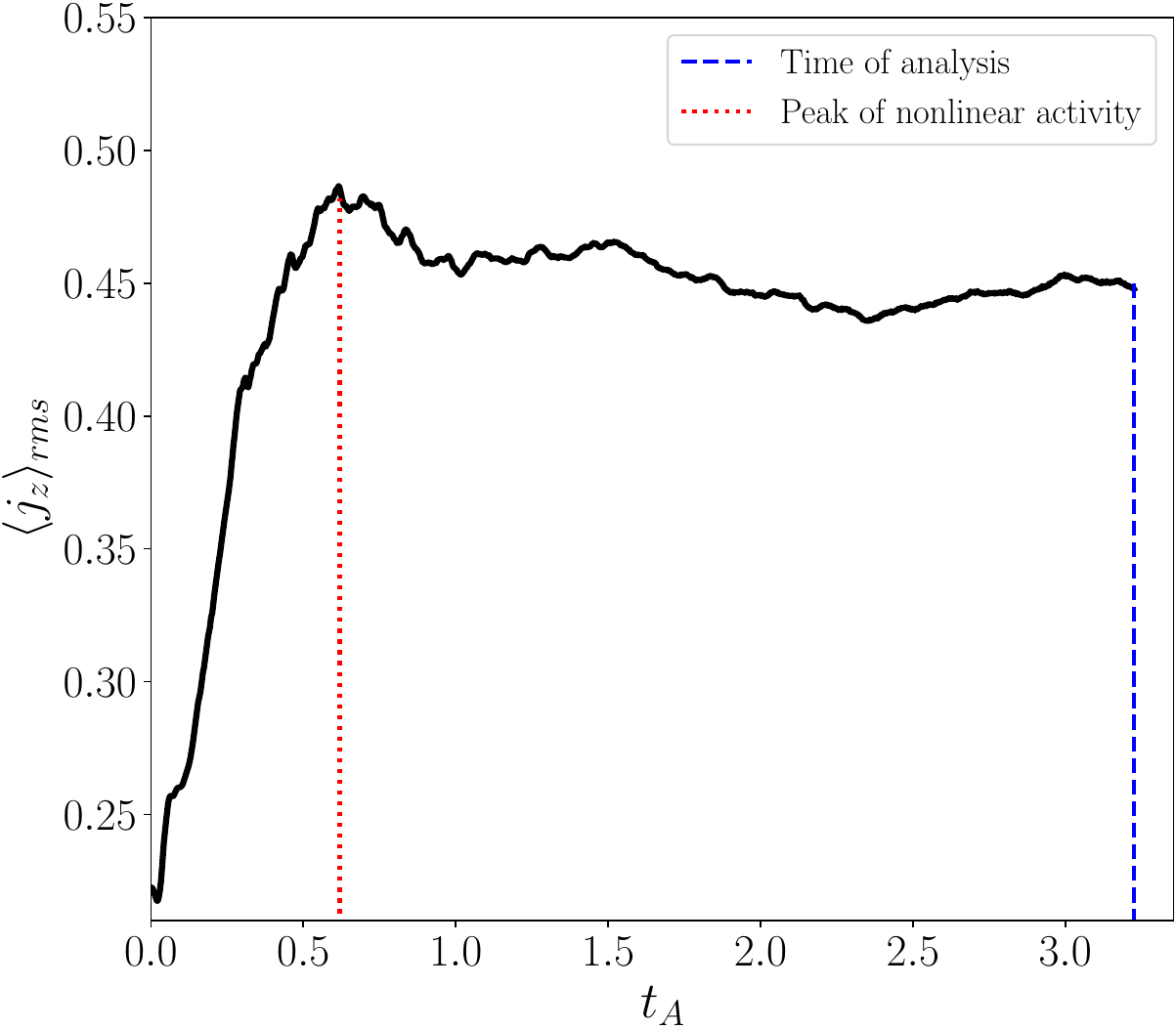}
	\caption{Rms of the vertical component of the current density (black solid line). Dotted (red) and dashed (blue) lines represent the time of most intense nonlinear activity and the analysis time, respectively.}
	\label{fig:Jz2}
\end{figure}

The setup settings of the simulation are described in the following. We use periodic boundary conditions and a spatial resolution such that the mesh is $\Delta x = \Delta y = d_e/3$, employing $10$ particles per cell (PPC), namely $5$ ions and $5$ electrons, with the full mass ratio. Time resolution holds $\Delta t = 0.45 \, \Delta x = 0.15 \, d_e$. The magnetization (which determines the available magnetic energy per particle) $\sigma := b_{0z}^2/4 \pi w_0$, where the out-of-plane mean magnetic field magnitude $b_{0z} \approx 43$  and the enthalpy density $w_0 := (n_p m_p + n_e m_e)c^2 + \Gamma_e \varrho_e + \Gamma_p \varrho_p$, with the internal energy density $\varrho_{\alpha} \approx n_{\alpha} k_B T_{\alpha}(\Gamma_{\alpha} - 1)^{-1}$ and the adiabatic index $\Gamma_{\alpha} = 4/3$. Since we are in a weakly relativistic regime, the magnetization $\sigma$ is approximately equal to $\sigma_p^{\text{cold}} := b_{0z}^2/4 \pi n_p m_p c^2$. The choice $\sigma \approx \sigma_p^{\text{cold}} = 1$ indicates an equilibrium between magnetic and kinetic forces, which is typical in astrophysical scenarios such as black hole winds near accretion disks.  Additionally, the plasma beta is $\beta_p = \beta_e := 8 \pi n_0 k_B T_{\alpha}/b_{0z}^2 = 3 \times 10^{-3}$, where the temperature $T_{\alpha} := \theta_{\alpha} m_{\alpha} c^2/k_B$ is chosen to ensure initial thermal equilibrium between species. We note that the choice of both $\sigma$ and $\beta$ is crucial for determining the regime of plasma \citep{ball2018electron}. The adimensional temperatures for each species are equal to $\theta_p = 1.5 \times 10^{-3}$ and $\theta_e \approx 2.75$, respectively. Such a configuration allows us to largely resolve the Debye plasma length $\lambda_D := \sqrt{(m_e \theta_e + m_p \theta_p) c^2/4 \pi n_0 e^2} \approx 2.35 d_e$. Besides, when the turbulence is fully developed, the velocity distribution of electrons is highly non-thermal, and their Larmor radius becomes significantly larger due to huge accelerations, effectively increasing our resolution. We further define the Alfv\'enic crossing time as $t_A := L_0/v_A \approx 7748$, hence the Alfv\'en velocity reads $v_A := c \sqrt{\sigma/1 + \sigma} \approx 0.71$.

We impose large-scale, random initial conditions for the magnetic field power spectrum to achieve a strong turbulent state. To avoid excessive compressive activity, no out-of-plane magnetic field fluctuations are prescribed at the beginning and no bulk flows or density perturbations are initiated.

The analysis for our fiducial simulation is carried out at a time $t \approx 3.23 ~ t_A$ when the peak of nonlinear activity has long been reached. Indeed, as shown in Fig.~\ref{fig:Jz2}, the rms of the vertical component of the current density $j_z = \left( \bm{\nabla} \times \bm{b}/ 4 \pi \right) \cdot \hat{z}$ experiences an absolute maximum much earlier, around $t \approx 0.62 ~ t_A$.

\section{The Kinetic Equilibrium Principle} \label{sec:kineq}

As discussed in the \textit{Letter}, in line with the Harris equilibrium, we consider a particle distribution function with an exponential dependence on the three invariants, as represented in Eq.~(1). Following this ansatz, the probability density of finding a particle at a given point in the phase space can be explicitly written as
\begin{equation}
\begin{aligned}
& \!\!\! f_{\alpha} \left( r, v_r, v_{\phi}, v_z \right) \! = \! N_{0 \alpha} \! \left( \dfrac{m_{\alpha}}{2 \pi k_B T_{\alpha}} \right)^{3/2} \\
& \!\!\! \exp \bigg[ \! - \! \dfrac{m_{\alpha}}{2 k_B T_{\alpha}} (v_{r}^2 \! + \! v_{\phi}^2 \! + \! v_z^2) \! + \! \dfrac{m_{\alpha} v_{\alpha}^{*}}{k_B T_{\alpha}} v_z + \! \dfrac{m_{\alpha} \Omega_{\alpha}^{*}}{k_B T_{\alpha}} r v_{\phi} \\[2pt]
& \!\!\! + \! \dfrac{q_{\alpha} v_{\alpha}^{*}}{c \, k_B T_{\alpha}} A_z(r) \! + \! \dfrac{q_{\alpha} \Omega_{\alpha}^{*}}{c \, k_B T_{\alpha}} r A_{\phi}(r) \! - \! \dfrac{q_{\alpha}}{k_B T_{\alpha}} \psi(r) \! \bigg],
\end{aligned}
\label{eq:equilibrium2}
\end{equation}
which must satisfy the Maxwell's equations:
\begin{equation}
\begin{aligned}
& \!\!\! \bm{\nabla} \cdot \bm{E} \! = \! 4 \pi e \bigg( \! \int_{\bm{\Omega}_{\bm{v}}} \!\!\! f_p( r, \bm{v}) d^3 \bm{v} \! - \! \int_{\bm{\Omega}_{\bm{v}}} \!\!\! f_e( r, \bm{v}) d^3 \bm{v} \! \bigg), \\
& \!\!\! \bm{\nabla} \cdot \bm{B} \! = \! 0, \\ 
& \!\!\! \bm{\nabla} \times \bm{E} \! = \! - \! \dfrac{1}{c} \dfrac{\partial \bm{B}}{\partial t} \! = \! \bm{0}, \\
& \!\!\! \bm{\nabla} \times \bm{B} \! = \! \dfrac{4 \pi e}{c} \bigg( \! \int_{\bm{\Omega}_{\bm{v}}} \!\!\! \bm{v} f_p( r, \bm{v}) d^3 \bm{v} \! - \! \int_{\bm{\Omega}_{\bm{v}}} \!\!\! \bm{v} f_e( r, \bm{v}) d^3 \bm{v} \! \bigg).
\end{aligned} \label{eq:system}
\end{equation}
We point out we are assuming stationarity, i.e. $\partial/\partial t = 0$. Given that the fields depend only on the radial coordinate, thus $\partial/\partial \phi = \partial/\partial z = 0$, Eq.s (\ref{eq:system}) reduce to three ODEs for the potentials (one for the electric potential and two for the azimuthal and vertical components of the vector potential): 
\begin{equation}
\begin{aligned}
& \!\!\!\!\!\! \dfrac{1}{r} \dfrac{d}{d r} \left( \! r \dfrac{d \psi}{d r} \right) \! = \! - 4 \pi e \bigg( \! \int_{\bm{\Omega}_{\bm{v}}} \!\!\! f_p d^3 \bm{v} \! - \!\! \int_{\bm{\Omega}_{\bm{v}}} \!\!\! f_e d^3 \bm{v} \! \bigg), \\
& \!\!\!\!\!\! \dfrac{d}{d r} \left( \dfrac{1}{r} \dfrac{d (r A_{\phi})}{d r} \right) \! = \! - \dfrac{4 \pi e}{c} \bigg( \! \int_{\bm{\Omega}_{\bm{v}}} \!\!\! v_{\phi} f_p d^3 \bm{v} \! - \!\! \int_{\bm{\Omega}_{\bm{v}}} \!\!\! v_{\phi} f_e ~ d^3 \bm{v} \! \bigg), \\
& \!\!\!\!\!\! \dfrac{1}{r} \dfrac{d}{d r} \left( \! r \dfrac{d A_z}{d r} \right) \! = \! - \dfrac{4 \pi e}{c} \bigg( \! \int_{\bm{\Omega}_{\bm{v}}} \!\!\! v_z f_p d^3 \bm{v} \! - \!\! \int_{\bm{\Omega}_{\bm{v}}} \!\!\! v_z f_e d^3 \bm{v} \! \bigg).
\end{aligned}
\label{sistema}
\end{equation} 
By taking the moments of Eq.~(1), it is possible to compute the particle density $N_{\alpha}(r) = \exp{ \left[ \gamma_{\alpha}(r) \right] }$, where $\gamma_{\alpha}(r)$ is given by Eq.~(2), and the bulk velocities $U_{\alpha z}(r) = v_{\alpha z}^*$ and $U_{\alpha \phi}(r) = \Omega_{\alpha}^* r$. The physical meaning of the free parameters $v_e^*$ and $\Omega_e^*$ is specified in Sec. \ref{sec:kinmodel} of the leading part of the current work, where the kinetic model is illustrated. Having these quantities, the current densities $J_{\phi \alpha}(r) = q_{\alpha} \, N_{\alpha}(r) \, U_{\phi \alpha}(r)$ and $J_{z \alpha}(r) = q_{\alpha} \, N_{\alpha}(r) \, U_{z \alpha}(r)$ are derived straightforwardly.

We observe a small and approximately constant net charge separation, which we define as
\begin{equation}
\kappa = N_e(r)- N_i(r) = \exp{ \left[\gamma_e(r)\right]} - \exp{\left[\gamma_p(r)\right]}, 
\label{net}
\end{equation} 
where $\kappa$ is a constant; hence, the calculation can be further simplified by assuming ${\bm \nabla} \cdot {\bm E} = - \nabla^2 \psi = - 4 \pi e \kappa$. This assumption directly leads to a straightforward choice for the electric potential of the form $\psi(r) = \psi(0) + \left( \pi e \kappa \right) r^2$. Due to gauge invariance, we set all potentials to zero at the vortex axis, i.e., $\psi(0) = 0$, $A_z(0) = 0$, and $A_{\phi}(0) = 0$. Introducing the quantity $\eta(r) = \gamma_e(r)-\gamma_p(r)$, Eq.~(\ref{net}) becomes 
\begin{equation}
\kappa =\exp{ \left[\gamma_e(r)\right]}- \exp{\left[\gamma_e(r) - \eta(r)\right]}, 
\label{net2}
\end{equation} 
which immediately yields $\eta(r) = - \ln{\{1 - \kappa \exp{ \left[- \gamma_e(r) \right]}\}}$. The relationship between the exponents $\gamma_e(r)$ and $\gamma_p(r)$ leads to
\begin{equation}
\begin{aligned}
& \!\!\!\!\!\! C_p \! + \! \dfrac{q_p \, v_p^{*}}{c \, k_B T_p} A_z(r) \! + \! \dfrac{q_p \, \Omega_p^{*}}{c \, k_B T_p} r A_{\phi}(r) \! - \! \dfrac{q_p}{k_B T_p} \psi(r) \\
& \!\!\!\!\!\! + \! \dfrac{m_p}{2 \, k_B T_p} \Omega_p^{{*}^2} r^2 \! + \! \dfrac{m_p}{2 k_B T_p} v_p^{{*}^2} \! =  C_e \! + \! \dfrac{q_e \, v_e^{*}}{c \, k_B T_e} A_z(r) \! \\
& \!\!\!\!\!\! + \! \dfrac{q_e \, \Omega_e^{*}}{c \, k_B T_e} r A_{\phi}(r) \! - \! \dfrac{q_e}{k_B T_e} \psi(r) \! + \! \dfrac{m_e}{2 \, k_B T_e} \Omega_e^{{*}^2} r^2 \\[2pt]
& \!\!\!\!\!\! + \! \dfrac{m_e}{2 \, k_B T_e} v_e^{{*}^2} - \eta(r),
\end{aligned} \label{eq:exp}
\end{equation}
where $\eta(r)$ is a small function due to the minimal net charge separation (consistent with quasi-neutrality) and $C_{\alpha} := \ln{(N_{0 \alpha})}$. By employing the correspondence between the exponents at the vortex center, where $\gamma_p(0) = \gamma_e(0) \, - \, \eta(0)$, we derive clear expressions for $C_p$ and $C_e$, which are then introduced into Eq.~(\ref{eq:exp}). Noticing that $\eta(r) - \eta(0)$ is a negligible term, we perform some algebra to obtain the following polynomial relation:
\begin{equation}
\zeta_0 \psi(r) + \zeta_1 A_z(r) + \zeta_2 A_\phi(r) ~ r + \zeta_3 r^2 = 0.
\end{equation}
A possible solution can be found by setting $\zeta_1 = 0$ and $\zeta_2 = 0$, from which the conditions shown in Eq.~(3) follow. The latter recall the Harris equilibrium in the sheet pinch for a non-homogeneous magnetic field in cylindrical geometry, with an additional condition for the in-plane azimuthal flow. Under this choice, a simple relation for the electric potential can be made explicit:
\begin{equation}
\psi(r) = - \frac{\zeta_3}{\zeta_0} r^2 \! = \! \left( \pi e \kappa \right)r^2 = \frac{ m_p {\Omega_p^{*}}^2 T_e - m_e {\Omega_e^{*}}^2 T_p}{2 e (T_p + T_e)} r^2.
\end{equation}
After that, we invoke Eq.~(\ref{eq:equilibrium2}) in the last two differential equations of the system~(\ref{sistema}) to finally arrive at the set of governing Eq.s~(4)--(7) of the main paper, which represents the core of the KVR model.

\section{Data-driven optimization model} \label{sec:data-driven}

\begin{figure}[b]
	\centering
	\includegraphics[width=0.99\columnwidth]{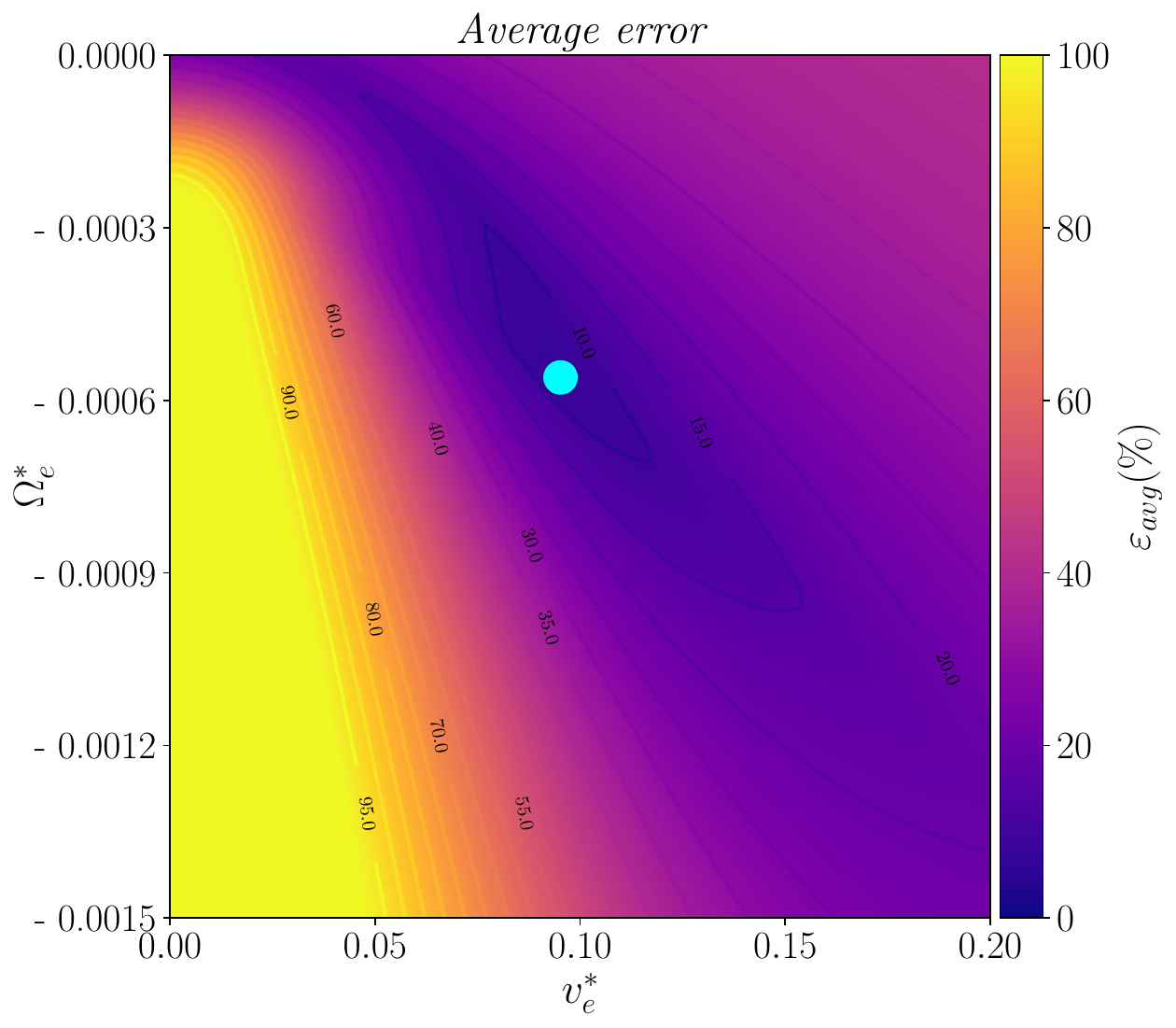}
	\caption{Average error estimated by considering all the vortex profiles. The best choice for the free parameters is highlighted by a (cyan) circle.}
	\label{fig:errALL}
\end{figure}

The KVR system of equations can be solved through direct numerical integration using a second-order Runge-Kutta scheme once the typical bulk electron velocity $v_{e}^*$ and angular velocity $\Omega_{e}^*$ are given, and boundary conditions at each vortex center are applied. For each coherent structure, we search for this pair of free parameters by implementing a data-driven, Monte Carlo numerical technique capable of selecting the best fit for each vortex, starting from the results of the integration on a parameter space $\bm{\mathcal{S}}$ given by $\{v_e^{*}, ~ \Omega_e^{*}\}$. 

Inspecting the most representative vortex in Fig.~\ref{fig:speccontours}-(b) at a time beyond the maximum of nonlinear activity ($t \approx 3.23 ~ t_A$), we impose $B'_z(0) \approx 40.5$, $B_{\phi}(r)/r |_{r = 0} = 0$, and $A_{\phi}(r)/r |_{r = 0} = 0$ as boundary conditions. As for the species temperatures ($T_e, ~ T_p$), we choose the averaged parallel temperature profiles in the vortex eye, i.e. $T_{\alpha} = T_{\parallel \alpha}$, with $T_{\parallel e} = 51.8$ and $T_{\parallel p} = 51.7$. 

Each long-lived structure is uniquely determined by the choice of the star parameters $v_e^{*}$ and $\Omega_e^{*}$. Such a choice has to ensure that the distribution function, once integrated over the phase space $\bm{\Omega}$, returns the different fields and momenta in the best possible agreement with the data. Hence, a Monte Carlo method is performed over the parameter space $\bm{\mathcal{S}}$, calculating the discrepancies between the KVR outputs and the observed profiles as below.

We span the parameter space by discretizing the domain $\{v_e^{*}, ~ \Omega_e^* \} \in \{[-0.2, 0.2], ~ [-0.0015, 0.0015] \}$, performing a campaign of reconstructions to minimize the error
\begin{equation}
\varepsilon_g(v_e^{*}, ~ \Omega_e^{*}) \! := \! \frac{\int_0^{r_0} \left[g_{\mathrm{KVR}}(r) \! - \! g_{\mathrm{sim}}(r)\right]^2 dr}{\int_0^{r_0} g_{\mathrm{sim}}(r)^2 dr}, 
\end{equation}
where $g_{\mathrm{sim}}$ is a generic field from the simulation and $g_{\mathrm{KVR}}$ is the reconstructed field from Eq.s~(4)--(7). We minimize for $g = [A_{\phi}, ~ A_z, ~ B_{\phi}, ~ B'_z, ~ J_{\phi}, ~ J_z, ~ N_e, ~ N_p]$, attaining for each of them the associated error $\varepsilon_g$. To achieve the best possible profiles, we estimate a single average over all the fields, namely $\varepsilon_{avg} (v_e^*, ~ \Omega_e^*) \! = \bar{\varepsilon}_g (v_e^*, ~ \Omega_e^*)$.

In Fig.~\ref{fig:errALL}, we show a zoom of the average error $\varepsilon_{avg}$, with its minimum located in the fourth quadrant of the parameter space (positive $v_e^*$ and negative $\Omega_e^*$). The minimum is placed at $v_e^{*} = 0.09$ and $\Omega_e^{*} = - 5.30 \times 10^{-4}$, as indicated by a (cyan) circle.

The resulting parameters are closely related to the flux tube current density: for the KVR model we get $\bm{J}^*(r) := q_e\left[ 0, ~ \left( N_p(r) \, \Omega_p^* - N_e(r) \, \Omega_e^* \right)r, ~ \left( N_p(r) \, v_p^* - N_e(r) \, v_e^* \right) \right]$, from which $\langle J^*_{\phi}\rangle_r \approx 1.17 \times 10^{-2}$ and $\langle J^*_z\rangle_r \approx -2.73 \times 10^{-2}$; for the observed structures in Fig.~\ref{fig:speccontours}-(b), we measure $\langle J_{\phi}\rangle_r \approx 1.27 \times 10^{-2}$ and $\langle J_z\rangle_r \approx -4.69 \times 10^{-2}$. Such an estimate indicates a very good agreement between the model and the data.

\section{Momentum equation and hierarchy of forces} \label{sec:momentum}

Computing the moments of the Vlasov-Maxwell system, one finds the momentum equation, which, in a stationary case, reads 
\begin{equation}
\bm{\nabla} \cdot \left( \rho_m \bm{U} \otimes \bm{U} \right) + \bm{\nabla} \cdot \bm{P}_{tot} - \rho_c \bm{E} - \bm{J} \times \bm{B} = \bm{0}, \label{Euler1}
\end{equation}
where $\rho_m$ represents the total mass density, $\rho_c$ the total charge density, and $\bm{P}_{tot}$ the total pressure tensor. Employing the continuity equation in the above expression, recalling the divergence of a symmetric tensor of rank two in cylindrical coordinates, and imposing $U_r = 0, ~ \partial/\partial \phi = \partial/\partial z = 0$, one can decompose Eq. (\ref{Euler1}) along the three cylindrical axes:
\begin{eqnarray}
&& \!\!\!\!\!\!\!\! 
\begin{aligned} \hat{r}: & \! - \! \dfrac{\rho_m \, U_{\phi} U_{\phi}}{r} \! + \! \left( \dfrac{1}{r} \partial_r (r \, P_{r r}) \! - \! \dfrac{P_{\phi \phi}}{r} \right) \! - \! \rho_c E_r \\
& - \! \left( J_{\phi} B_z - J_{z} B_{\phi} \right) \! = \! 0; 
\end{aligned} \label{eq:mom_1} \\
&& \!\!\!\!\!\!\!\! \hat{\phi}: \! \left( \! \partial_r P_{r \phi} \! + \! \dfrac{2 P_{r \phi}}{r} \! \right) \! - \! \rho_c E_{\phi} \! - \! \left( J_z B_r \! - \!  J_r B_z \right) \! = \! 0; \label{eq:mom_2} \\
&& \!\!\!\!\!\!\!\! \hat{z}: \! \dfrac{1}{r} \partial_r (r \, P_{r z}) \! - \! \rho_c E_z \! - \! \left( J_r B_{\phi} \! - \! J_{\phi} B_r \right) \! = \! 0. \label{eq:mom_3}
\end{eqnarray}
Since the radial components of currents and magnetic field are null, the out-of-diagonal terms of the pressure are negligible, and the main electric field is radial, it is worth noting that the only nontrivial equation is the radial one.

\begin{figure}[ht]
	\centering
	\includegraphics[width=0.99\columnwidth]{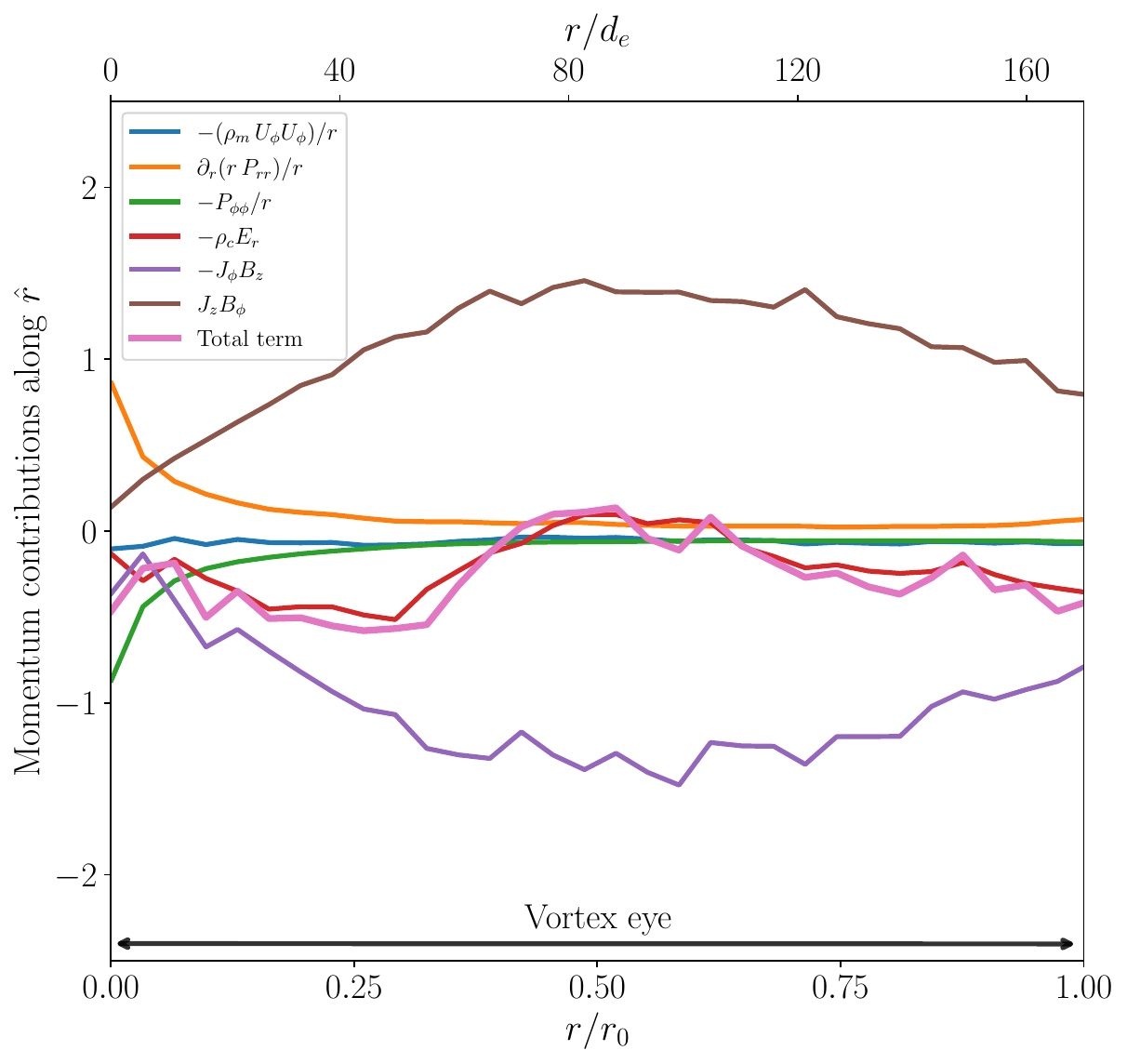}
	\caption{Components of the momentum equation for the fluid approximation in the vortex eye.}
	\label{fig:modelkvr2}
\end{figure}

Establishing a hierarchical order for all the terms appearing in the radial momentum equation proves helpful in extracting the critical constituents of such metastable states. We compute all the quantities of Eq.~(\ref{eq:mom_1}) and compare them in Fig. \ref{fig:modelkvr2}. Computing radial averages for each of them, it is evident that magnetic forces dominate the dynamics, in fact, a clear ordering can be inferred: $|J_i B_j|_{i,j \not = r} > |\rho_c E_r| \gg |\left( {\bm \nabla} \cdot \bm{P}_{tot} \right) \cdot \hat{\bm{r}}|>|\left[ {\bm \nabla} \cdot \left(\rho_m ~ \bm{U} \otimes \bm{U} \right) \right] \cdot \hat{\bm{r}}|$.


\bibliography{BIBLIO_IMBROGNO}
\bibliographystyle{aasjournal}



\end{document}